\documentclass[12pt,showpacs,amsmath,amssymb]{revtex4}
\usepackage{graphicx}

\begin{document}

\renewcommand{\thesection}{\arabic{section}}
\renewcommand{\thesubsection}{\arabic{section}.\arabic{subsection}}
\renewcommand{\theequation}{\arabic{section}.\arabic{equation}}

\title{{Hypothetical first order $|\Delta S|=2$ transitions \\ in the $K^{0}-\overline{K^{0}}$
 complex }}
\author{  J. Piskorski}
\email{jaropis@proton.if.uz.zgora.pl}
\affiliation{University of Zielona G\'ora, Institute of Physics, \\
Plac Slowianski 6, 65-069 Zielona Gora, Poland.}

\begin{abstract}
The influence of a hypothetical $CP$ violating $|\Delta S|=2$
interaction on the time  evolution of the $K^{0} -
\overline{K^{0}}$ system is investigated. It is shown, that if we
were to assume the existence of the superweak-like interaction
then this would lead to the conclusion, that there might be
observable effects in the masses of the neutral kaons. We address the possibility of experimental observation of these effects and perform a computer simulation of one of the parameters which describe such effects. Instead of the widely used Lee, Oehme and Yang approximation which is not suitable to considering this kind of interaction we use a formalism based on the Kr\'olikowski-Rzewuski equation.
\end{abstract}
\maketitle

\section{Introducton}
Until recently there were two types of models of interactions which were
considered plausible while investigating the source of the CP
violation \cite{1}. The miliweak models which assume that a part
of order $10^{-3}G_{F}$ in the weak interaction was responsible
for the observed CP violation effects. One of the most important
predictions of this class of models is that the CP violation
should be observed also in other than
$K^{0}\rightleftarrows\overline{K^{0}}$ processes, and that it
should be of the same order. The CKM model is an
example of such miliweak models, at the same time being the most
successful one. The recent experimental results concerning the
measurement of $\frac{\epsilon'}{\epsilon}$ and the $CP$ violation
in the neutral $B$-meson system show, that the CKM model correctly
describes the $CP$ violation. There is, however, a small
possibility that a superweak-like interaction does exist, and some
authors consider its implications (see \cite{2} and references
therein). 

In the present paper we consider an effect which would be present,
if the superweak interaction, or an interaction of a similar
nature (the terms 'superweak' and 'superweak-like' will be used interchangeably), really existed in nature in addition to the CKM mechanism.
We find, that the standard Weisskopf-Wigner approach to studying
the $K^{0}\rightleftarrows\overline{K^{0}}$ process is not
sensitive enough for the study of such an interaction, and
therefore we choose the Krolikowski-Rzewuski approach
\cite{3}-\cite{4} and its extensions to the neutral kaon system
suggested in \cite{5}-\cite{8}. The paper is organized as
follows. In the second section we review the most important (for
our purposes) features of the Standard Model and the Superweak
model and their present experimental status. The third section
reviews the standard phenomenological approach to the neutral kaon
system, which is based on the Weisskopf-Wigner approach. Also, basing on \cite{9}, \cite{5}-\cite{8}, we review an alternative formalism and analyze its relevance to the
superweak-like interaction. The fourth section contains a computer
simulation of the time dependence of one of the parameters
introduced in the alternative model, namely the difference of the
diagonal elements of the effective Hamiltonian, which in the
presence of the superweak interaction turns out to be different
from zero. The summary and conclusions are contained in the last
section.
\section{$K^{0} \rightleftarrows \overline{K^{0}}$  mixing in the Standard Model and the Superweak Model}
In this section we quickly review the Standard Model approach to
the $K^{0} \rightleftarrows \overline{K^{0}}$. We also briefly
describe the salient features of the superweak scenario of $CP$
violation.
\subsection{$K^{0}\rightleftarrows \overline{K^{0}}$  mixing and CP violation in the Standard
Model}

\begin{figure}
\includegraphics{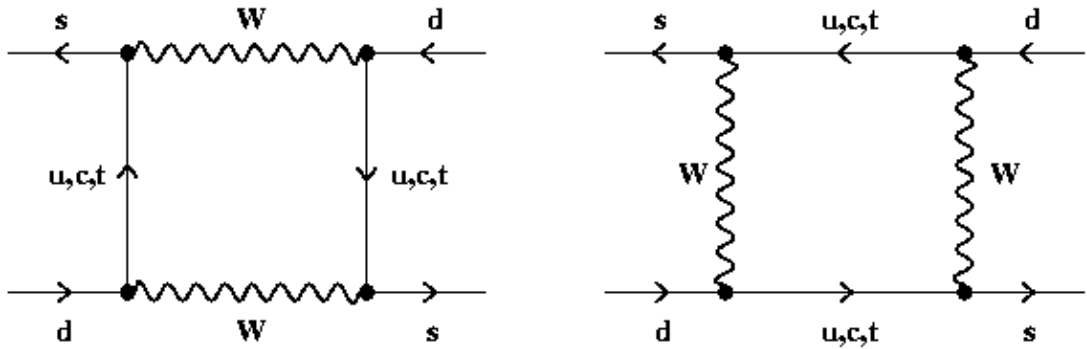}
\caption{\label{fig:box} Diagrams corresponding to the $K^{0}
\rightleftarrows \overline{K^{0}}$ process.}
\end{figure}

The flavour transitions allowed in the Standard Model are
specified by the CKM matrix, which allows the following flavour
mixing:
\begin{equation}
\left(\begin{array}{c}d'\\s'\\b'\end{array}\right)=
\left(\begin{array}{ccc}
 V_{ud}&V_{us}&V_{ub}\\
 V_{cd}&V_{cs}&V_{cb}\\
 V_{td}&V_{ts}&V_{tb}\\
 \end{array}\right)
\left(\begin{array}{c}d\\s\\b\end{array}\right) \label{CKM}.
\end{equation}
Consequently, to the lowest order transitions
$K^{0}\rightleftarrows\overline{K^{0}}$ can proceed through the
diagrams presented in figure FIG.~\ref{fig:box}.

In the CKM theory there are no direct, first order
$K^{0}\rightleftarrows\overline{K^{0}}$ transitions. In other
words there are no first order $|\Delta S|=2$ transitions, or, in
yet another equivalent formulation, which we will be using in the
remaining part of the paper:
\begin{equation}
\langle{\bf 1}|H^{(1)}|{\bf 2}\rangle=0,\end{equation}
where $|K_{0}\rangle \equiv|{\bf 1}\rangle$ and $| {\overline
K}_{0}\rangle \equiv |{\bf 2}\rangle$ and $H^{(1)}$ is the
flavour-changing part of the weak Hamiltonian \cite{10}.

Matrix \ref{CKM} is unitary and contains $9$ parameters. Three of
these parameters may be chosen to be real angles $\theta_{12}$,
$\theta_{13}$, $\theta_{23}$ and the remaining six are phases. The
number of phases can be reduced by using the fact, that spinors
are defined up to a phase, so we may redefine the quark
eigenstates. After doing this we notice, that in the procedure
there are only five independent phase differences, whereas there
are six phases in \ref{CKM}, so there is one physically meaningful
phase in this unitary matrix. This is the crucial point of the
$CKM$ theory because this phase allows for $CP$ violation \cite{1}.
\setcounter{equation}{0}
\subsection{The hypothetical
Superweak interaction}  The Superweak model postulates the existence of a
new $|\Delta S|=2$ interaction, which violates CP. The coupling
constant of this interaction should be smaller than second order
weak interaction. Thus, the superweak model assumes a
non-vanishing first order transition matrix element:
\begin{equation}
\langle{\bf 1}|H^{(1)}|{\bf 2}\rangle\sim gG_{F}\neq 0,
\end{equation}
where $g$ is the superweak coupling constant. 
It is widely accepted that this interaction can only be detected
in the $K_{L} - K_{S}$, because it is the only known pair of
states with the energy difference so small, that it is sensitive
to interactions weaker than second order weak interaction \cite{1}.

\subsection{The status of the Standard Model and the Superweak
model}

 The recent experimental results from the CPLEAR and KTeV
Collaborations and others have given the decisive answer to the
question whether the CP violation effects are correctly described
by the $CKM$ miliweak theory. The measured value of
$\frac{\epsilon'}{\epsilon}=(1.72\pm0.18)\times 10^{-3}$ \cite{2}
proves that there is a direct CP violating effect, and that CP
violation cannot only be  ascribed to mass mixing in the
$K^{0}\rightleftarrows\overline{K^{0}}$ process. On the contrary:
the CKM must be the dominant source of $CP$ violation (in
low-energy flavour-changing processes)\cite{2}. Additionally, the
measured value is perfectly consistent with the world average for
the value $\frac{\epsilon'}{\epsilon}$  \cite{11}. Another experimental argument for the miliweak CKM theory are the two recent measurements of $CP$
violation in $B$ decays (\cite{2} and references therein). In
other words, the Standard Model alone is able to correctly predict
the value of $\frac{\epsilon'}{\epsilon}$ and no improvements or
extensions are in fact necessary.

However, even if the CP violation effects are described by the CKM
the the idea of a $|\Delta S|=2$ interaction has not been
abandoned entirely. Indeed, some authors consider the implications
of such an interaction - the question of the existence of the
superweak interaction turns out to be of some importance in the "tagged"
experiments in which the flavour is determined for the initial
meson and then the for the meson at the time of decay. The
existence of the $|\Delta S|=2$ superweak interaction might cause
the production of the "wrong" neutral meson states \cite{12}. The
effect of this hypothetical interaction is believed to be
negligibly small as it would add to the
$K^{0}\rightleftarrows\overline{K^{0}}$ mixing. 

In the remaining part of the paper we want to discuss the implications of this
hypothetical interaction for the time evolution of the
$K^{0}-\overline{K^{0}}$ complex in the case of $CPT$
conservation in view of the recent experimental facts. We show that if we want to assume the existence of
such an interaction then we can no longer use the LOY
approximation. We also address the question of how this kind of interaction
could have observable effects in the decay of the neutral mesons. The size of the effects introduced by the new, hypothetical interaction is also investigated, with the assumption that the recent experimental results and their interpretation are correct.  To achieve this goal we will use a different approach than the LOY approximation, namely, a formalism based on the Kr\'olikowski-Rzewuski equation \cite{3}, \cite{4}. 
\setcounter{equation}{0}
\section{The standard phenomenological description of the
 $K^{0}-\overline{K^{0}}$ system}
 In this section we briefly describe the phenomenology which is
 currently used to describe the time evolution of the
 $K^{0}-\overline{K^{0}}$ system. 
\subsection{The Lee, Oehme and Yang (LOY) approximation}
This formalism is based on the formalism of
 particle mixture introduced by Gell-Mann and Pais \cite{11a}. The most
 important modification to this formalism was introduced by Lee,
 Oehme and Yang \cite{10a}, who, using the Weisskopf-Wigner approximation
 arrived at the formula (\ref{mixingmatrix}) - see below - which is currently used.
 Further extensions were introduced by many other authors, e.g. Bell and Steinberger \cite{12a}.

 In the standard approach the full Hamiltonian is divided into two
 parts:
 \begin{equation}
 H=H^{(0)}+H^{(1)},
 \label{hamiltonian}
 \end{equation}
 where $H^{(0)}$ is the flavour-conserving part of the
 Hamiltonian, and $H^{(1)}$ is the flavour-changing part. The
complete state vector which has evolved from ${\bf |1\rangle}$ or
${\bf |2\rangle}$ is projected onto the subspace spanned by ${\bf
|1\rangle}$ and ${\bf |2\rangle}$. Therefore we define the state
vector as:
\begin{equation}
|\Psi(t)\rangle=\alpha_{1} (t) {\bf |1\rangle}+\alpha_{2} (t) {\bf
|2\rangle}.
\end{equation}
Lee, Oehme and Yang, by modifying the Weisskopf-Wigner method for
the single line, showed that the time dependence of the vector
$\left(\begin{array}{c}\alpha(t)\\\beta(t)\end{array}\right)$ can
be described by the following Schr\"odinger-like equation:
\begin{equation}
i\frac{d}{dt}\left(\begin{array}{c}\alpha_{1}(t)\\\alpha_{2}(t)\end{array}\right)=
\left(\begin{array}{ll}H_{11}^{LOY}&H_{12}^{LOY}\\H_{21}^{LOY}&H_{22}^{LOY}\end{array}\right)
\left(\begin{array}{c}\alpha_{1}(t)\\\alpha_{2}(t)\end{array}\right),\end{equation}
where we have adopted $\hbar=c=1$ and the matrix elements are
matrix elements of the weak interaction transition operator. In
the case of $CPT$ conserved it can be shown, that for this
effective Hamiltonian  we have $H_{11}^{LOY}=H_{22}^{LOY}$, but we have no
information on $H_{12}^{LOY}$ or $H_{21}^{LOY}$ \cite{10}.

The effective Hamiltonian can be split into two parts, each of
them with a definite physical meaning, namely:
\begin{equation}
H^{LOY}=M-i\Gamma/2,
\label{hloy}
\end{equation}
or
\begin{equation}
i\frac{d}{dt}\left(\begin{array}{c}\alpha_{1}(t)\\\alpha_{2}(t)\end{array}\right)=
\left(\begin{array}{cc}M-i\Gamma/2&M_{12}-i\Gamma_{12}/2\\M_{21}-i\Gamma_{21}/2&M-i\Gamma/2
\end{array}\right)
\left(\begin{array}{c}\alpha_{1}(t)\\\alpha_{2}(t)\end{array}\right)\label{mixingmatrix}.\end{equation}
The diagonal entries bear no indices, as we want to further
emphasize the main result of the LOY approach, namely $H_{11}^{LOY}=H_{22}^{LOY}$.

For our purposes, which is the analysis of the influence of the
hypothetical $|\Delta S|=2$ interaction on the time development of
the $K^{0}-\overline{K^{0}}$ system, the LOY method is not
suitable. Indeed, in \cite{14}, \cite{8}  it was shown, that the
LOY formulae may only be correct if we assume $\langle{\bf
1}|H^{(1)}|{\bf 2}\rangle=0$ and take $t\rightarrow\infty$. This obviously excludes the possibility of using the Lee, Oehme and Yang  approximation in studying a hypothetical superweak interaction

\subsection{The alternative approach } 

One alternative  to the approach described above is the formalism developed in
\cite{5}---\cite{8}. We will briefly review this approximation
and its basic findings.

The starting point of the derivation of an alternative effective
Hamiltonian carried out in \cite{8,9,14} is the Kr\'olikowski-Rzewuski
Equation \cite{3,4}. In this approach the time evolution is not
studied in the total space of states $\cal{H}$ but rather in a
closed subspace $\cal{H}_{||}\subset\cal{H}$. If we define the following
projector:

\[P\stackrel{\rm def}{=} |\bf{1}\rangle \langle \bf{1}|+|\bf{2}\rangle \langle \bf{2}|,\]
then the subspace $\cal{H}_{||}$ may be defined as $\cal{H}_{||} =
P {\cal H}$ or $| \psi ;t\rangle_{||} = P| \psi ;t\rangle$.
 In this way the total state space is split into two orthogonal
subspaces $\cal{H}_{\parallel}$ and
$\cal{H}_{\perp}=\cal{H}\ominus\cal{H}_{\parallel}$, and the
Shr\"odinger equation can be replaced by equations describing each
of the  subspaces respectively. The equation for $\cal{H}_{\parallel}$ has the following form
\cite{3,4,9}:
\begin{equation}
\left(i\frac{\partial}{\partial t}-PHP\right)|\psi;t\rangle_{||}=|
\chi ;t\rangle-
i\int_{0}^{\infty}K(t-\tau)|\psi;\tau\rangle_{||}d\tau,
\label{evolutparal}
\end{equation}
\begin{equation}
Q=I-P,
\end{equation}

\begin{equation}
K(t)=\Theta(t)PHQe^{-iQHQ}QHP,
\end{equation}
\begin{equation}
|\chi;t\rangle=PHQe^{-iQHQ}|\psi\rangle_{\perp},
\end{equation}
where
\[
\Theta(t)=\left\{
\begin{array}{lll}
1&{\rm for}&t\geq0\\ 0&{\rm for}&t<0 \end{array} \right. .\]
Following \cite{3,4} we introduce an effective Hamiltonian:
\begin{equation}
H_{||}(t)\equiv PHP + V_{||}(t).
\end{equation}
This formula corresponds to (\ref{hloy}), which also specifies an effective Hamiltonian.

Now, the main difference between the standard Lee, Oehme and Yang approximation and this approach is the effective potential. It can be shown that \cite{8,9}:
\begin{equation}
V_{||}(t)\simeq V_{||}^{1}(t)=-i\int_{0}^{\infty}K(t-\tau)e^{i(t-\tau)PHP}P d\tau.
\label{Vpara}
\end{equation}
To establish notation let us now define the following symbols:
\begin{equation}
PHP\equiv\left[\begin{array}{ll} H_{11}&H_{12}\\H_{21}&H_{22}\end{array}\right]
,\end{equation}
\[H_{ij}=\langle {\bf j}|H|{\bf i}\rangle\verb+ +, H_{0}\equiv\frac{1}{2}(H_{11}+H_{22}), \verb+ + \kappa\equiv\sqrt{|H_{12}|^{2}+\frac{1}{4}(H_{11}-H_{22})}, \verb+ + H_{z}\equiv\frac{1}{2}(H_{11}-H_{22}).
\]

The matrix elements $v_{ij}(t)=\langle {\bf j}|V_{||}(t)|{\bf i \rangle}$ of  $V_{||}$ (\ref{Vpara}), without assuming any symmetries, like $[CP,H]=0$ or $[CPT,H]=0$ \cite{8,9} are:
\begin{eqnarray}
v_{j1}(t) = & - & \frac{1}{2} \Big( 1 + \frac{H_{z}}{{\kappa}}
\Big){\Xi}_{j1} ( H_{0} + {\kappa},t )
\label{v-j1-imp1} \\
& - & \frac{1}{2} \Big( 1 - \frac{H_{z}}{{\kappa}}
\Big){\Xi}_{j1} (
H_{0} - {\kappa},t )  \nonumber  \\
& - & \frac{H_{21}}{2 {\kappa}} {\Xi}_{j2} (H_{0} +
{\kappa},t)+ \frac{H_{21}}{2 {\kappa}} {\Xi}_{j2} (H_{0} -
{\kappa},t ) , \nonumber
\end{eqnarray}
\begin{eqnarray}
v_{j2}(t) = & - & \frac{1}{2} \Big( 1 - \frac{H_{z}}{{\kappa}}
\Big){\Xi}_{j2} (H_{0} + {\kappa},t)
\label{v-j2-imp1} \\
& - & \frac{1}{2} \Big( 1 + \frac{H_{z}}{{\kappa}}
\Big){\Xi}_{j2} (
H_{0} - {\kappa},t )  \nonumber  \\
& - & \frac{H_{12}}{2 {\kappa}} {\Xi}_{j1} (H_{0} + {\kappa},t
) + \frac{H_{12}}{2 {\kappa}} {\Xi}_{j1} (H_{0} - {\kappa},t )
, \nonumber
\end{eqnarray}

where
\begin{equation}
\Xi(\lambda,t)\stackrel{\rm def} {=} PHQ \frac{e^{-it(QHQ -
\lambda)}-1}{QHQ -\lambda}QHP,
\end{equation}

and ${\Xi}_{jk} ( \varepsilon,t ) = \langle  {\bf j} \mid \Xi (
\varepsilon,t ) \mid{\it {\bf k}} \rangle$, and $j,k =1,2$. 
This effective potential, together with the remaining parts
of the effective Hamiltonian yields the following matrix elements
for the effective Hamiltonian:
\begin{equation}
h_{jk}(t) = \langle{\bf j}|H_{||}|{\bf k}\rangle = H_{jk} +
v_{jk}(t), \; \; ({\scriptstyle j,k =1,2})\label{fullhamiltonian}
.\end{equation}
For the $[CPT,H]=0$ case the formulae simplify as $H_{z}=0$ in this case

Now, it is easy to notice that, in the case of $[CPT,H]=0$, contrary to the LOY effective Hamiltonian for which we have $H_{11}^{LOY}-H_{22}^{LOY}=0$ , the difference between the diagonal elements is non-vanishing:

\begin{equation}
h_{z}(t)=\frac{1}{2}(h_{11}(t)-h_{22}(t))\neq 0, (t>0). 
\label{inequal} 
\end{equation} 
It is also obvious that the necessary condition for (\ref{inequal}) to be true is $H_{12}\neq0$, that is, the existence a superweak
interaction.
\setcounter{equation}{0}
\section{Computer simulation of the time evolution $h_{z}(t)$ within
the Friedrichs-Lee model}

In this section we perform a numerical simulation of the $h_{z}(t)$
parameter, which has proved so important in the present approach.
By making some assumptions concerning the scale of the
hypothetical superweak interaction we arrive at a form which is
convenient for computer analysis. We also provide figures
demonstrating the time evolution of the module and real and
imaginary part of this parameter.
\subsection{The Friedrichs-Lee model}
In \cite{8} the Friedrichs-Lee model was used to obtain the
following formulae for the matrix elements of the effective
Hamiltonian with the assumption $[CPT,H]=0$:
\begin{equation}
h_{j1}(t)=m_{j1}-\frac{1}{2}\left(
\Gamma_{j1}+\frac{m_{21}}{|m_{12}|}\Gamma_{j2}\right)
\Phi_{0}(t;m_{0}+|m_{12}|-\mu)-\label{hajeden}\end{equation}
\[-\frac{1}{2}\left(\Gamma_{j1}-\frac{m_{21}}{|m_{12}|}
\Gamma_{j2}\right)\Phi_{0}(t;m_{0}-|m_{12}|-\mu),
\]
\begin{equation}
h_{j2}(t)=m_{j2}-\frac{1}{2}\left(
\Gamma_{j2}+\frac{m_{12}}{|m_{12}|}\Gamma_{j1}\right)
\Phi_{0}(t;m_{0}+|m_{12}|-\mu)-\label{hadwa}\end{equation}
\[-\frac{1}{2}\left(\Gamma_{j2}-\frac{m_{12}}{|m_{12}|}
\Gamma_{j1}\right)\Phi_{0}(t;m_{0}-|m_{12}|-\mu).
\]
In these formulae $m_{0}\equiv \langle {\bf 1}|H^{(0)}|{\bf
1}\rangle=\langle {\bf 2}|H^{(0)}|{\bf 2}\rangle$, compare
Eq.(\ref{hamiltonian}), $m_{12}\equiv H_{12}$; $m_{0}-\mu$ is the
difference between the mass of the mesons considered and the
threshold energy of the continuum state, like $K\rightarrow 2\pi$.
 Functions  $\Phi_{0}(t,m)$ are defined by:
 \begin{equation}
\Phi_{0}(t,m)=F_{0}(m)-F_{0}(t,m),\end{equation} where
\begin{equation}
F_{0}(t,m)=\frac{a}{\sqrt{m}}[S(\sqrt{mt})-C(\sqrt{mt})]-
\end{equation}
\[-i\frac{a}{\sqrt{m}}[C(\sqrt{mt})+S(\sqrt{mt})-1],\]
\begin{equation}
F_{0}(m)=i\frac{a}{\sqrt{m}},
a=(m_{11}-\mu)^{\frac{1}{2}},
\end{equation}
and finally $S(x)$ and $C(x)$ are the sine and cosine Fresnel
integrals:
\[C(x)=\frac{1}{\sqrt{2}}\int_{0}^{x^{2}}\frac{\cos(\tau)}{\sqrt{\tau}}d\tau,\]
\[S(x)=\frac{1}{\sqrt{2}}\int_{0}^{x^{2}}\frac{\sin(\tau)}{\sqrt{\tau}}d\tau.\]
 The parameters $\Gamma_{ij}$ correspond
to the matrix elements of the decay matrix in the LOY
approximation (\ref{mixingmatrix}). 

By setting $j=1$ in
(\ref{hajeden}) and $j=2$ in (\ref{hadwa}) and then substracting
(\ref{hadwa}) from (\ref{hajeden})we get:
\begin{equation}
h_{z}(t)=\frac{m_{21}\Gamma_{12}-m_{12}\Gamma_{21}}{4|m_{12}|}[\Phi_{0}(t;m_{0}-|m_{12}|-\mu)-
\Phi_{0}(t;m_{0}+|m_{12}|-\mu)].
\end{equation}
Using a new independent variable $x$, defined in Appendix A we can rewrite $h_{z}(t)$ as:
\begin{equation}
h_{z}(x(t))=\frac{m_{21}\Gamma_{12}-m_{12}\Gamma_{21}}{4|m_{12}|}r(x),
\end{equation} 
where $r(x)$ has the following form (see Appendix A):
\pagebreak
 \begin{equation}
 r(x)=\left(1-\frac{|m_{12}|}{m_{0}-\mu}\right)^{-\frac{1}{2}}\left\{i-
 \left[S\left(\sqrt{\left(1-\frac{|m_{12}|}{m_{0}-\mu}\right)x}\right)-C\left(\sqrt{\left(1-\frac{|m_{12}|}{m_{0}-\mu}\right)x}\right)\right]
 \right. \label{pelnadelta}
 \end{equation}
\[\left.-i\left[C\left(\sqrt{\left(1-\frac{|m_{12}|}{m_{0}-\mu}\right)x}\right)+S\left(\sqrt{\left(1-\frac{|m_{12}|}{m_{0}-\mu}\right)x}\right)\right]
\right\}\]
 \[
+ \left(1+\frac{|m_{12}|}{m_{0}-\mu}\right)^{-\frac{1}{2}}\left\{-i+
 \left[S\left(\sqrt{\left(1+\frac{|m_{12}|}{m_{0}-\mu}\right)x}\right)-C\left(\sqrt{\left(1+\frac{|m_{12}|}{m_{0}-\mu}\right)x}\right)\right]
 \right.
\]
\[\left.-i\left[C\left(\sqrt{\left(1+\frac{|m_{12}|}{m_{0}-\mu}\right)x}\right)+S\left(\sqrt{\left(1+\frac{|m_{12}|}{m_{0}-\mu}\right)x}\right)\right]
\right\}.\]

In spite of its complicated appearance, expression
(\ref{pelnadelta}) is simple to analyze using computer methods as
it contains no other variables but the independent variable $x$.

To extract any numerical information from (\ref{pelnadelta})  we need
to make some assumptions concerning the strength of the superweak
interaction. There are some estimates in the literature - we will
accept the one suggested by Lee in \cite{10} (equation 15.138, page
375): $\frac{|m_{12}|}{m_{0}-\mu}\simeq 10^{-17}$. To be sure, we
do not even know if the strength is different from zero - we are
assuming {\it a} value of $\frac{|m_{12}|}{m_{0}-\mu}$ which is
consistent with the assumptions made in section 2. to see how
$h_{z}(t)$ changes with time.

\subsection{Time dependence of $h_{z}(x(t))$}
Below we present three figures: the time dependence of
$|r(x)|$, $\Re(r(x))$ and $\Im(r(x))$ - where $\Re$ and $\Im$ stand for the real and imaginary parts, respectively - which are
proportional to the corresponding parameters connected with
$h_{z}$, because from (\ref{defhz}), (\ref{defdelta}) we have (see Appendix A):
\begin{equation}
|r(x)|=\frac{|h_{z}(x(t))|}{|\Gamma_{12}|\cdot|2\sin(\phi-\theta)|}
\label{delta}.
\end{equation}

The figures correspond to the value of $\frac{|m_{12}|}{m_{0}-\mu}$=$10^{-17}$.

\begin{figure}
\includegraphics{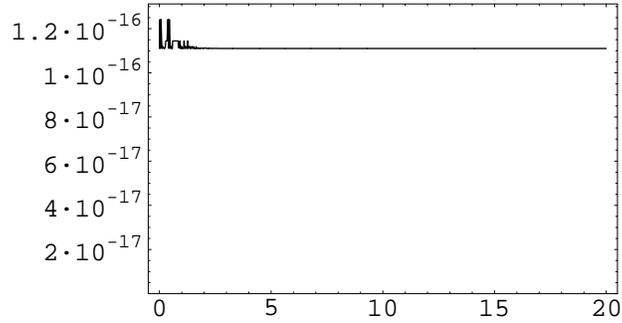}
\caption{\label{fig:log17} {The change of $y\sim|h_{z}(x)|$ in the
logarithmic scale for the value
$\frac{|m_{12}|}{m_{0}-\mu}=10^{-17}$.}}
\end{figure}

\begin{figure}
\includegraphics{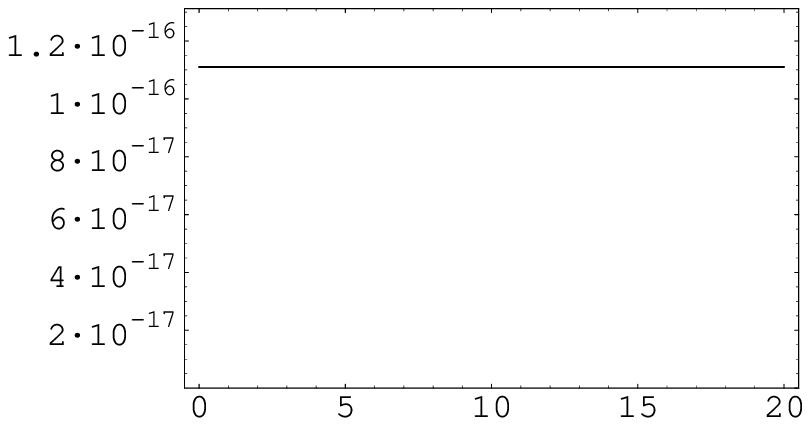}
\caption{\label{fig:rehz17} {The change of $y \sim \Re [h_{z}(x)]$ for
long times in the logarithmic scale for $\frac{|m_{12}|}{m_{0}-\mu}=10^{-17}$.}}
\end{figure}

\begin{figure}
\includegraphics{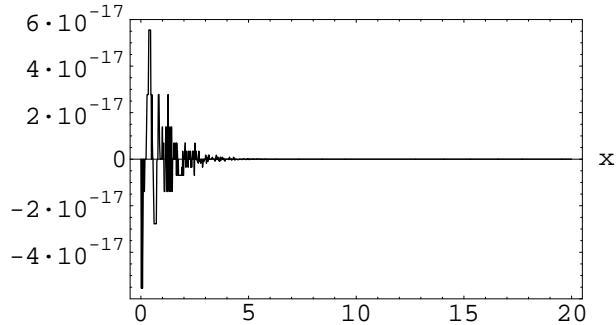}
\caption{\label{fig:imhz17} {The change of $y\sim \Im[h_{z}(t)]$ for
long times  in the logarithmic scale for $\frac{|m_{12}|}{m_{0}-\mu}=10^{-17}$.}}
\end{figure}

 In FIG.~\ref{fig:log17} we
show the time dependence of the module of $r(x)$, which is
basically the same as $|h_{z}(x)|$ This figure is in the
logarithmic scale - we can see, that the parameter $r(x)$
oscillates around the constant value of around $10^{-16}$.The
imaginary part rapidly tends to zero. Therefore
the real part of $h_{z}(x)$ is responsible for the non-vanishing of
$|h_{z}(x)|$ - this is clearly seen in FIG~\ref{fig:rehz17}. 

In the standard approach the real parts of the diagonal
elements of the effective Hamiltonian are interpreted as the
masses of the particles. Therefore it seems that the existence of
the superweak interaction would remove the mass degeneracy between
the particle and antiparticle in the neutral kaon system.
Correspondingly, the imaginary parts are interpreted as the decay
constants, so in the model considered the decay widths of the
particle and antiparticle should be equal, which is consistent
with the standard result. These results are consistent with the conclusions reached earlier on the basis of the form of $h_{z}(t)$ for large times - see Appendix B. 

\subsection{Order-of-magnitude estimation of the effect introduced by the superweak interaction}
In this short subsection we  try to estimate the order of magnitude of the effect introduced by the hypothetical superweak-like interaction. To this end we use the assumption, that the dominating contribution to $|\Gamma_{12}|$ is correctly described by the Standard Model. This means that  we may assume 
\begin{equation}
|\Gamma_{12}|\sim\frac{G^{2}_{F}M_{P}^{4}\sin^{2}\theta}{(2\pi)^{4}}m_{K_{0}}\sim
10^{-12}MeV,
\end{equation}
where $G_{F}$ is the Fermi constant, $M_{P}$ is the
proton mass and $\theta$ is the Cabbio angle \cite{15}. Using our result form the previous section, $r(x\rightarrow \infty)\sim 10^{-16}$ and Equation \ref{delta} we get the
following upper bound on our  parameter: 
\begin{equation}
\frac{|h_{z}(t\rightarrow \infty)|}{m_{K_{0}}}\lesssim
\frac{1}{2}10^{-31}.
\end{equation}
This value corresponds to the presently measured $|\frac{m_{K_{0}}-m_{\overline{K_{0}}}}{m_{K_{0}}}|\leq10^{-18}$
\cite{16}. Obviously, the effect calculated in the present paper is much too small to be observed with the present, and possibly also future, experiments.
 With this result, the question of the utility of the very concept of the superweak interaction arises.

\section{Summary and Conclusions}
In the present paper the influence of the hypothetical superweak
interaction on the time evolution of the neutral kaon complex has
been studied. We have investigated the dependence of the effective
Hamiltonian on the very existence of such an interaction and
found, that the standard Weisskopf-Wigner approach, used for
example by Lee, Oehme and Yang \cite{10} may only be correct if we
assume the nonexistence of this interaction. This is the reason
for choosing another approach, namely the formalism based on the
Krolikowski-Rzewuski equation \cite{5}-\cite{8}, which makes no
such assumptions and is sensitive to the existence of the first
order $|\Delta S|=2$ interaction. The two approaches have been
shown to be equivalent in e.g. \cite{14} in the absence of this interaction, but its inclusion leads to a prediction which is different from the
generally accepted result. 

The computer simulation of the time dependence of the parameter
$|h_{z}(t)|$, corresponding to the difference of the diagonal elements of the effective Hamiltonian,  presented in section 5. shows, that this parameter is
different from zero and basically constant. Another result of this
section is that it is the real part which is responsible for the
non-vanishing $|h_{z}(t)|$, as the imaginary part very rapidly tends to zero. All these results would be impossible to reach without the
Krolikowski-Rzewuski approach, further developed in
\cite{5}-\cite{8}, as the standard method due to Lee, Oehme and
Yang is insensitive to the superweak interaction.

The  prediction of the standard Lee-Oehme and Yang states that the real parts of the diagonal elements of the effective Hamiltonian governing the time evolution of the
neutral kaon system should be equal. In the usual interpretation, the real parts of the diagonal elements correspond to the masses of the particles and hence the masses of $K^{0}$ and $\overline{K^{0}}$ are equal. In the approach assumed in the present paper the real parts of the effective halimtonian are equal on condition that there is no superweak interaction.  The result of the present paper, and many others \cite{5}-\cite{10}, \cite{9}, \cite{10}, is that mathematically the difference between the masses of particles and antiparticles needs not be equal to zero, if we assume the standard identification of the diagonal elements of the effective Hamiltonian plus the existence of the superweak-like interaction. However, even if this mathematical property of the effective Hamiltonian were to produce some results which are meaningful for the experiment, sensitivity of the instruments would have exceed the present day experiments by a few orders of magnitude (see section 4.3). 
Naturally, the question of the utility of the theory developed
above should be raised. It is quite possible, that this formalism and the numerical results
may only be used to confirm the standard result laid out in
\cite{10a}. 

Finally, it should be stressed that all the results obtained in
the present paper are consistent with the Standard Model and the
recent experimental findings. We have been assuming, that even if
there is a CP violating $|\Delta S|=2$ interaction, the $K^{0} -
\overline{K^{0}}$ complex is correctly described by the Standard
Model to a high degree of accuracy. This is the reason for
assuming $\frac{|m_{12}|}{m_{0}-\mu}=10^{-17}$ in section 5. and
$|\Gamma_{12}|\sim 10^{-12}MeV$ in section 4.3.
\begin{acknowledgments}
I would like to thank professor Krzysztof Urbanowski for many helpful discussions.
\end{acknowledgments}
\appendix*
\section{A}
By rewriting the $m_{12}$ and $\Gamma_{12}$ parameters as:
\begin{equation}
m_{21}=m_{12}^{*}\equiv |m_{12}|e^{-i\theta},
\end{equation}
\[\Gamma_{12}=\Gamma_{21}^{*}\equiv |\Gamma_{12}|e^{i\phi},\]
we may cast $h_{z}(t)$ in the following form:
\begin{equation}
h_{z}(t)=\frac{1}{4}|\Gamma_{12}|(e^{i(\phi-\theta)}-e^{-i(\phi-\theta)})\times
\label{defhz}\end{equation}
\[\times[\Phi_{0}(t;m_{0}-|m_{12}|-\mu)-
\Phi_{0}(t;m_{0}+|m_{12}|-\mu)] =\]
\[=i|\Gamma_{12}|2\sin(\phi-\theta)[\Phi_{0}(t;m_{0}-|m_{12}|-\mu)-
\Phi_{0}(t;m_{0}+|m_{12}|-\mu)].\]

It is easy to notice, that if $\phi=\theta$ we have $h_{z}=0$, but
this case corresponds exactly to the $CP$ conserved case - compare
\cite{8} page 3743. Let us assume from now on, that we are
dealing with the $CP$-violating, $CPT$-conserving case in which
$\sin(\phi-\theta)\neq 0$.

 To make our formulae simpler, let us define:
\begin{equation}
r(t)=[\Phi_{0}(t;m_{0}-|m_{12}|-\mu)-
\Phi_{0}(t;m_{0}+|m_{12}|-\mu)] \label{defdelta}.\end{equation}
 So now
\begin{equation}
r(t)=F_{0}(m_{0}-|m_{12}|-\mu)-F_{0}(t;m_{0}-|m_{12}|-\mu)-\label{appdelta}\end{equation}
\[-F_{0}(m_{0}+|m_{12}|-\mu)+F_{0}(t;m_{0}+|m_{12}|-\mu).\]
Let us now transform the above expression using:
\[\sqrt{m_{0}-|m_{12}|-\mu}=\sqrt{m_{0}-\mu}\cdot\sqrt{1-\frac{|m_{12}|}{m_{0}-\mu}},\]
and
\[\sqrt{m_{0}+|m_{12}|-\mu}=\sqrt{m_{0}-\mu}\cdot\sqrt{1+\frac{|m_{12}|}{m_{0}-\mu}},\]
and a new, dimensionless variable
\begin{equation}
x\equiv(m_{0}-\mu)\cdot t.
\end{equation}
If we define $\tau_{L}=5.2\cdot 10^{-8}s$ as the mean lifetime of $K_{L}$, the value of $x$ corresponding to this time is $x\simeq 2\cdot 10^{16}$. 

Now we have:
\begin{equation}(m_{0}-|m_{12}|-\mu)t=(m_{0}-\mu)(1-\frac{|m_{12}|}{m_{0}-\mu})t=(1-\frac{|m_{12}|}{m_{0}-\mu})x,\end{equation}
\[(m_{0}+|m_{12}|-\mu)t=(m_{0}-\mu)(1+\frac{|m_{12}|}{m_{0}-\mu})t=(1+\frac{|m_{12}|}{m_{0}-\mu})x.\]
Using the above and Equation (\ref{appdelta}) we arrive at (\ref{pelnadelta}). 

\section{B}
In \cite{8} the following formula for the behaviour of $h_{z}(t)$ in the long time region was derived:
\begin{eqnarray}
h_{z}(t) \simeq_{t\rightarrow \infty}\frac{m_{21}\Gamma_{12}-m_{12}\Gamma_{21}}{4|m_{12}|}\left\{F_{0}(m_0-|m_{12}|-\mu)-F_{0}(m_0-|m_{12}|-\mu)\right.
\\ \frac{1+i}{\sqrt{2\pi}}\frac{a_{1}(0)}{\sqrt{t}}e^{-i(m_{0}-\mu)t}\left[\frac{e^{i|m_{12}|t}}{m_{0}-|m_{12}|-\mu}-\frac{e^{-i|m_{12}|t}}{m_{0}+|m_{12}|-\mu}\right]\left.\right\}.\nonumber
\end{eqnarray}
Let us write out the real and imaginary parts of the above expression in terms of masses, $\Gamma_{ij}$ and $|m_{ij}|$:
\begin{eqnarray}
\Re (h_{z}(t\rightarrow\infty))=-(m_{11}-\mu)^{\frac{1}{2}}\frac{\Im(m_{21}\Gamma_{12})}{2|m_{12}|}\left\{\frac{\sqrt{m_{0}-\mu+|m_{12}|}-\sqrt{m_{0}-\mu-|m_{12}|}}{\sqrt{(m_{0}-\mu)^{2}-|m_{12}|^{2}}}+\right.\\
+\frac{1}{\sqrt{2\pi}\sqrt{t}}\left[\frac{\sin(m_{0}-\mu-|m_{12}|)t-\cos(m_{0}-\mu-|m_{12}|)t}{m_{0}-\mu-|m_{12}|} \right.\nonumber\\ \left.\left.-\frac{\sin(m_{0}-\mu+|m_{12}|)t-\cos(m_{0}-\mu+|m_{12}|)t}{m_{0}-\mu+|m_{12}|}\right]\right\}, \nonumber
\end{eqnarray}
\begin{eqnarray*}
\Im(h_{z}(t\rightarrow \infty))=-(m_{11}-\mu)^{\frac{1}{2}}\frac{1}{\sqrt{2\pi}\sqrt{t}}\frac{\Im(m_{21}\Gamma_{12})}{2|m_{12}|}\left\{\frac{\sin(m_{0}-\mu-|m_{12}|)t+\cos(m_{0}-\mu-|m_{12}|)t}{m_{0}-\mu-|m_{12}|}\right.\\-\left.\frac{\sin(m_{0}-\mu+|m_{12}|)t+\cos(m_{0}-\mu+|m_{12}|)t}{m_{0}-\mu+|m_{12}|}\right\}.
\end{eqnarray*}
From these formulae it is clear, that the imaginary part tends to zero and the real part tends to a constant.

\pagebreak

\begin{thebibliography}{99}
\bibitem{1}K.Kleinknecht, $CP$ Violation in the $K^{0} - \overline{K}^0$ System in $CP$ Violation, Eds. C.Jarlskog, World Scientific, Singapore, (1989).
\bibitem{2}Y.Nir, $CP$ Violation - A New Era, Lectures given at the 55th Scottish Universities Summer School in Physics , Heavy Flavour Physics, 2001.
\bibitem{3} W.Kr\'olikowski and J.Rzewuski, Bull. Acad. Polon. Sci. 4,19 (1956).
\bibitem{4} W.Kr\'olikowski and J.Rzewuski, Nuovo Cimento {\bf B 25}, 739 (1975), and references therein.
\bibitem{5} K.Urbanowski, Phys. Lett. {\bf A171}, (1992) 151.
\bibitem{6} K.Urbanowski, Int. J. Mod. Phys. {\bf A9} (1994).
\bibitem{7} K.Urbanowski, Phys. Rev. {\bf A 50}, (1994) 2847.
\bibitem{8} K.Urbanowski, Int. J. Mod. Phys. {\bf A 8}, (1993) 3721.
\bibitem{9} K.Urbanowski, J.Piskorski, Found.Phys., Vol.30, No. 6,2000.
\bibitem{10} T.D.Lee, Particle Physics and Introduction to Field Theory, Harwood Academic Publishers GmbH, Chur, Switzerland, (1990).
\bibitem{10a} T.D. Lee, R.Oehme and C.N.Yang, Phys. Rev., {\bf 106}, (1957), 340.
\bibitem{11} J.Ellis, N.E. Mavromatos, Phys.Rept.320:341-354,1999.
\bibitem{11a} M.Gell-Mann, A.Pais, Phys.Rev.{\bf 97}, 1387 (1955). 
\bibitem{12} L.Lavoura, P.J.Silva, Phys.Rev.D60:056003,1999.
\bibitem{12a} J. S. Bell and J. Steinberger, Proceedings, Oxford International Conference on Elementary Particles, (1965). 
\bibitem{14} J.Piskorski, Acta Phys. Polon., Vol. 31 (2000).
\bibitem{15} G.D'Ambrosio, G.Isidori, Int.J.Mod.Phys. {\bf A13},(1998).
\bibitem{16} D.E.Groom et al., The European Physical Journal {\bf C15} (2000).


\end{thebibliography}
\end{document}